\documentclass{article}
\usepackage{amsmath}  
\usepackage{empheq}   
\usepackage{graphicx}
\usepackage{subfigure}
\usepackage{float}
\usepackage{adjustbox} 
\usepackage{booktabs}  
\usepackage{epstopdf}
\usepackage{amsmath,amsthm,amsfonts,amssymb}
\usepackage{xcolor}
\usepackage[a4paper, total={6in, 8in}]{geometry}

\def\opone{\leavevmode\hbox{\small1\kern-3.8pt\normalsize1}}

\newcommand{\beq}{\begin{equation}}
\newcommand{\eeq}{\end{equation}}
\newcommand{\beqa}{\begin{eqnarray}}
\newcommand{\eeqa}{\end{eqnarray}}

\newcommand{\al}{\alpha}
\newcommand{\be}{\beta}
\newcommand{\om}{\Omega}

\newcommand{\br}{\mathbb{ R}}

\newcommand{\ome}{\omega}

\newcommand{\ga}{\gamma}

\newcommand{\La}{\Lambda}
\newcommand{\te}{\theta}

\newcommand{\la}{\lambda}

\newcommand{\ep}{\varepsilon}

\makeatletter
\renewcommand{\@fnsymbol}[1]{\ifcase#1\or 1\or 2\or 3\or 4\or 5\or 6\or 7\or 8\or 9\else\@ctrerr\fi}
\makeatother

\title{Modeling the Impact of Misinformation Dynamics on Antimicrobial Resistance}
\author{
Laurance Fakih\thanks{Corresponding author: \texttt{laurance.fakih@upb.ro}}, 
	Andrei Halanay\thanks{\texttt{andrei.halanay@upb.ro}} \\
	\small Department of Mathematics and Informatics, University Politehnica of Bucharest, Romania
}
\date{}

\begin{document}
	\maketitle
	
	\footnotetext[1]{Corresponding author: \texttt{laurance.fakih@upb.ro}}
	\footnotetext[2]{\texttt{andrei.halanay@upb.ro}}

\begin{center}
	\textit{This is a preliminary version. An extended version with numerical results is forthcoming}
\end{center}

\begin{abstract}
Antimicrobial Resistance (RAM) poses a significant threat to global public health, making important medicines less useful. While the medical and biological reasons behind RAM are well studied, we still don't know enough about how false health information affects people's actions, which can speed up RAM. This study presents a new mathematical model to investigate the complex interplay between the spread of misinformation and the dynamics of RAM. We adapt a multi-strain fake news model, including distinct population compartments representing individuals susceptible to, believing in, or skeptical of various ideas related to antibiotic use. The model considers multiple "strains" of misinformation, such as the wrong belief that antibiotics are effective for viral infections or not trusting medical advice regarding prudent antibiotic prescription. Time delays are integrated to reflect the latency in information processing, behavioral change, and the manifestation of resistance. Through stability analysis and numerical simulations, this research aims to identify critical factors and parameters that influence the propagation of harmful beliefs and their consequent impact on behaviors contributing to RAM. The findings could help develop public health campaigns to reduce the negative impact of misinformation on fighting antimicrobial resistance.\\
In this paper, we extend a fake news spreading model with time delays by adding two biological equations that capture antimicrobial resistance (RAM) dynamics. The original model divides the population into six compartments based on information exposure and belief status. We append two new variables to track the prevalence of antimicrobial resistance and inappropriate antibiotic consumption. The stability of the extended system's equilibria is analyzed using rigorous mathematical methods. We establish critical thresholds that determine whether misinformation spreads, whether it persists, and how resistance develops. Our theoretical results suggest that controlling misinformation can effectively reduce antimicrobial resistance development. Numerical simulations illustrate our analytical findings and provide insights into potential intervention strategies.
\end{abstract}
\section{Introduction}
With improvements in digital technology and social media platforms, the speed at which fake news spread has reached unprecedented levels.
The explosion of social networks has changed the way information is shared,~making it possible for real and false information to reach large groups of people. This development raises serious concerns about the accuracy of online content and the potential dangers of misinformation.
\\
While not all of them cause immediate harm, many have the potential to cause panic, distort public perception, and destroy one's reputation.
There is a critical necessity to investigate how fake news spread in order to develop effective strategies for fighting disinformation and creating an informed society.
\\
Antimicrobial resistance (RAM) represents one of the most significant public health challenges of our time. The World Health Organization has identified it as a global health emergency, with projections suggesting that by 2050, RAM could cause 10 million deaths annually if left unchecked \cite{oneill2016tackling}. While biological and pharmacological factors contribute significantly to RAM development, the role of human behavior—particularly as influenced by information and misinformation—has received increasing attention in recent years.

The spread of misinformation about antibiotics (e.g., "antibiotics are effective against viral infections" or "it's acceptable to stop taking antibiotics when symptoms improve") can lead to inappropriate use patterns that accelerate resistance development. Conversely, accurate information can promote appropriate antibiotic stewardship. This complex interplay between information dynamics and biological processes necessitates mathematical models that can capture both aspects and their interactions.

Many mathematical models have been proposed to study either misinformation spread or antimicrobial resistance separately. However, models that integrate both phenomena are relatively scarce. In this paper, we extend an existing model of fake news spreading with time delays by adding two new equations that capture the biological dynamics of antimicrobial resistance.

Our research provides a comprehensive theoretical analysis that aligns more closely with real-world dynamics of misinformation and antimicrobial resistance. By preserving the original structure of the fake news model and adding biological components, we create a framework that can be used to study the impact of information dynamics on public health outcomes.

\subsection{Overview of our model}
In this study, we denote by $(x_1, x_2, \ldots)$ the sub-population sizes (potential users, active users, skeptics, etc.), and we evaluate the equilibrium and stability conditions that determine rumor or fake news persistence.
\\
The original model describes the spread of two competing strains of fake news in a population. The population is divided into six compartments:
\begin{itemize}
	\item $x_1$: Susceptible/inactive users who have not yet engaged with the information
	\item $x_2$: Active users who are engaged with information but have not adopted any specific belief
	\item $x_3$: Believers of fake news strain 1
	\item $x_4$: Believers of fake news strain 2
	\item $x_5$: Skeptics/fact-checkers regarding fake news strain 1
	\item $x_6$: Skeptics/fact-checkers regarding fake news strain 2
\end{itemize}
The dynamics of these compartments are governed by the following system of delay differential equations:
\begin{empheq}[left=\empheqlbrace]{align}
	\dot{x}_{1} &= \Lambda - \mu x_{1} 	- \eta x_{1} x_{2\tau_{1}} 
	- \frac{\beta_{3} x_{1} x_{3}}{1 + \alpha_{3} x_{1} + \varepsilon_{3} x_{3}}- \frac{\beta_{4} x_{1} x_{4}}{1 + \alpha_{4} x_{1} + \varepsilon_{4} x_{4}}
	\label{eq:1}\\[5pt]
	\dot{x}_{2} &= \eta x_{1} x_{2\tau_{1}} - \mu x_{2}
	\label{eq:2}\\[5pt]
	\dot{x}_{3} &= \frac{\beta_{3} x_{1} x_{3}}{1 + \alpha_{3} x_{1} + \varepsilon_{3} x_{3}}- \mu_{3} \frac{x_{3\tau_{2}}}{1 + \nu_{3} x_{3\tau_{2}}}
	+ \gamma_{3} x_{5 \tau_{4}}	- \mu x_{3}	\label{eq:3}\\[5pt]
	\dot{x}_{4} &= \frac{\beta_{4} x_{1} x_{4}}{1 + \alpha_{4} x_{1} + \varepsilon_{4} x_{4}}- \mu_{4} \frac{x_{4\tau_{3}}}{1 + \nu_{4} x_{4\tau_{3}}}+ \gamma_{4} x_{6 \tau_{5}}- \mu x_{4}
	\label{eq:4}\\[5pt]
	\dot{x}_{5} &= \frac{\mu_{3} x_{3\tau_{2}}}{1 + \nu_{3} x_{3\tau_{2}}}- \gamma_{3} x_{5 \tau_{4}}- \mu x_{5}
	\label{eq:5}\\[5pt]
	\dot{x}_{6} &= \frac{\mu_{4} x_{4\tau_{3}}}{1 + \nu_{4} x_{4\tau_{3}}}- \gamma_{4} x_{6 \tau_{5}}- \mu x_{6}
	\label{eq:6}
\end{empheq}
where:
\begin{itemize}
	\item $\Lambda$ is the recruitment rate of susceptible users
	\item $\mu$ is the natural leaving/forgetting rate
	\item $\eta$ is the contact rate between active and inactive users
	\item $\beta_3, \beta_4$ are the transmission rates of fake news strains 1 and 2
	\item $\alpha_3, \alpha_4, \varepsilon_3, \varepsilon_4$ are saturation parameters for the Holling Type II incidence
	\item $\mu_3, \mu_4$ are the rates at which believers become skeptics
	\item $\nu_3, \nu_4$ are half-saturation constants for skeptic conversion
	\item $\gamma_3, \gamma_4$ are the rates at which skeptics relapse to believers
	\item $\tau_1, \tau_2, \tau_3, \tau_4, \tau_5$ are time delays in the system
\end{itemize}
Remark that $\dot{x_1}+\dot{x_2}+\dot{x_3}+\dot{x_4}+\dot{x_5}+\dot{x_6}=\La-\mu({x_1}+{x_2}+{x_3}+{x_4}+{x_5}+{x_6})$ and it is standard that  this implies the invariance of the set $$\om=\{(x_1,\dots,x_6)\in \br^6_{+}|{x_1}+{x_2}+{x_3}+{x_4}+{x_5}+{x_6}\leq \dfrac{\La}{\mu}\}$$
\subsection{Equilibria and basic calculations}
The fake-news-free equilibrium (FNFE) is $E_1$ with components  $$x_1=\dfrac{\La}{\mu},x_2=x_3=x_4=x_5=x_6=0.$$
Besides, the system has other equilibria: $E_2=(x_1^{*},x_2^{*},x_3^{*},x_4^{*},x_5^{*},x_6^{*})$, $ E_3=(\tilde{x_1},\tilde{x_2},\tilde{x_3}, 0,\tilde{x_5},0)$,  $E_4=(\hat{x_1},\hat{x_2},0,\hat{x_4}, 0,\hat{x_6})$,  $E_5=(x_1^*,0,x_3^*,x_4^*,x_5^*,x_6^*), E_6=(x_1^*,0,x_3^*,0*,x_5^*,0),E_7(x_1^*,0,0,x_4^*,0,x_6^*)$. Of course, the values of the components are specific for every equilibrium point but we  denote them again with $*$ to keep the notation less complicated.
\\
To calculate the components of $E_2$, remark first that, when $x_2\neq0$ one has $$x_1^*=\dfrac{\mu}{\eta}.$$ Then,  equations (5) and (6) give
$$x_5=\dfrac{\mu_3 x_3}{(1+\nu_3 x_3) (\mu+\ga_3)}, x_6=\dfrac{\mu_4 x_4}{(1+\nu_4 x_4) (\mu+\ga_4)}.$$ Since $x_3\neq 0$, equation (3) gives 
$$\be_3 x_1^* (\mu+\ga_3)(1+\nu_3 x_3)-\mu \mu_3 (1+\al_3 x_1^*+\ep_3 x_3)-\mu (\mu+\ga_3) (1+\nu_3 x_3)(1+\al_3 x_1^*+\ep_3 x_3)=0$$ that eventually gives a positive solution $x_3^*$. Introducing this solution in equation (1) one has
$$x_2^*=\dfrac{(\La-\mu x_1^*)(1+\al_3 x_1^*+\ep_3 x_3^*)-\be_3 x_1^* x_3^*}{\eta x_1^*(1+\al_3 x_1^*+\ep_3 x_3)}.$$ In order that $x_2^*>0$ one must have $\La-\mu x_1^*>0\Leftrightarrow\La-\dfrac{\mu^2}{\eta}>0$ and this implies, as will be seen in the next section, that $E_1$ is unstable. When $x_4\neq 0$, equation (4) gives 
$$\be_4 x_1^* (\mu+\ga_4)(1+\nu_4 x_4)-\mu \mu_4 (1+\al_4 x_1^*+\ep_4 x_3)-\mu (\mu+\ga_4) (1+\nu_4 x_4)(1+\al_4 x_1^*+\ep_4 x_4)=0$$ whose positive solution, if exists, will be $x_4^*$.
\section{Stability analysis for $E_1$}
Let $ E_1= (x_1^*,0, 0,0,0,0)$.where $x_1^{*}=\dfrac{\La}{\mu}$ without any relation with the same notation appearing in other equilibria.\\  I
The stability analysis of $E_1$ will be established through the use of a Lyapunov-Krasovskii functional, in a form inspired by \cite {RZ}. Let $v(x)=x-1-lnx$
\\
Define$$V(x_1,\dots,x_6, x_{2\tau_1})=x_1^{*} v\left(\dfrac{x_1}{x_1^{*}}\right)+x_2+x_3+x_4+x_5+x_6+\eta x_1^{*}\int_{-\tau_1}^0x_2(t+\te)d\te.$$ $V$ is positive definite in $\om\times C([-\tau_1,0], \br_{+})$ and is zero only for $E_1$. We show next that its derivative  is negative when some conditions for the parameters hold true.
$$
\begin{array}{l}
\dot{V}=v'\left(\dfrac{x_1^{*}}{x_1}\right)\dot{x_1}+\dot{x_2}+\dot{x_3}+\dot{x_4}+\dot{x_5}+\dot{x_6}+x_1^{*}\int_{-\tau_1}^0x_2(t+\te)d\te=\\=\left(1-\dfrac{x_1}{x_1^{*}}\right)\left(\La-\mu x_1-\eta x_1 x_{2\tau_1}-\be_3\dfrac{x_1 x_3}{1+\al_3 x_1+\ep_3 x_3}-\be_4\dfrac{x_1 x_4}{1+\al_4 x_1+\ep_4 x_4}\right)+\eta x_1 x_{2\tau_1}-\mu x_2+\\+\be_3\dfrac{x_1 x_3}{1+\al_3 x_1+\ep_3 x_3}-\mu_3\dfrac{x_{3\tau_2}}{1+\nu_3 x_{3\tau_2}}+\ga_3 x_{5\tau_4}-\mu x_3+\\ +\be_4\dfrac{x_1 x_4}{1+\al_4 x_1+\ep_4 x_4}-\mu_4\dfrac{x_{4\tau_3}}{1+\nu_4 x_{4\tau_3}}+\ga_4 x_{6\tau_5}-\mu x_4+\\ +\mu_3\dfrac{x_{3\tau_2}}{1+\nu_3 x_{3\tau_2}}-\ga_3 x_{5\tau_4}-\mu x_5+\mu_4\dfrac{x_{4\tau_3}}{1+\nu_4 x_{4\tau_3}}-\ga_4 x_{6\tau_5}-\mu x_6+\eta x_1^{*}x_2-\eta x_1^{*} x_{2\tau_1}=\\=-\dfrac{\mu )x_1^{*}-x_1)}{x_1}+\be_3\dfrac{x_1^{*} x_3}{1+\al_3 x_1+\ep_3 x_3}+\be_4\dfrac{x_1^{*} x_4}{1+\al_4 x_1+\ep_4 x_4}-\mu x_2-\mu x_3-\mu x_4-\mu x_5-\mu x_6+\eta x_1^{*}x_2\leq\\ \leq -\dfrac{\mu )x_1^{*}-x_1)}{x_1}+\be_3 x_1^{*} x_3-\mu x_3+\be_4 x_1^{*} x_4-\mu x_4+\eta x_1^{*}x_2-\mu x_2-\mu x_5-\mu x_6=\\=-\dfrac{\mu (x_1^{*}-x_1)}{x_1}+x_3 (\be_3 x_1^{*}-\mu)+x_4 (\be_4 x_1^{*}-\mu)+x_2 (\eta x_1^{*}-\mu)-\mu x_5-\mu x_6<0
\end{array}$$ 
if 
\begin{equation}
\La \be_3<\mu^2, \La \be_4<\mu^2, \La \eta<\mu^2.
\end{equation}
With the derivative of $V$ strictly negative in $\om$ and  $E_1$ the only zero of $V$, we conclude, following \cite{hale_lunel1993}, Theorem 5.3.1, that $E_1$ is globally asymptotic stable in $\om$.
\section{Characteristic equation for $E_2$}
The characteristic equation becomes:
\begin{equation}
\resizebox{\textwidth}{!}{%
	$
	\begin{vmatrix}
	\lambda -a_{11} &\eta x_1^{*} e^{-\la\tau_1}  & -a_{13} & -a_{14} & 0 & 0 \\[1mm]
	-\eta x_2^{*}& \lambda - a_{22}-b_{22}e^{-\la\tau_1} & 0 & 0 & 0 & 0 \\[1mm]
	-a_{31} & 0 & \lambda -a_{33}-c_{33}e^{-\la\tau_2} & 0 & -e_{35}e^{-\la\tau_4} & 0 \\[1mm]
	-a_{41} & 0 & 0 & \lambda -a_{44}-d_{44}e^{_\la\tau_3} & 0 & -f_{46}e^{-\la\tau_5}\\[1mm]
	0 & 0 & -c_{53}e^{-\la\tau_2}& 0 & \lambda -e_{55}e^{_\la\tau_4}-a_{55} & 0 \\[1mm]
	0 & 0 & 0& d_{64}e^{-\la\tau_5} & 0 & \lambda-f_{66}e^{-\la\tau_5}-a_{66} \\[1mm]
	
	\end{vmatrix} = 0
	$
}
\end{equation}
The presence of $a_{31}, a_{41}$ makes this equation difficult to handle. We apply the rank-one perturbation argument from the paper \cite{BBH}:if the modulus of these terms are small, in a sense made precise in \cite{BBH}, then, if the equation 
\begin{equation}
\resizebox{\textwidth}{!}{%
	$
	\begin{vmatrix}
	\lambda -a_{11} &\eta x_1^{*} e^{-\la\tau_1}  & -a_{13} & -a_{14} & 0 & 0 \\[1mm]
	-\eta x_2^{*}& \lambda - a_{22}-b_{22}e^{-\la\tau_1} & 0 & 0 & 0 & 0 \\[1mm]
	0 & 0 & \lambda -a_{33}-c_{33}e^{-\la\tau_2} & 0 & -e_{35}e^{-\la\tau_4} & 0 \\[1mm]
	0& 0 & 0 & \lambda -a_{44}-d_{44}e^{_\la\tau_3} & 0 & -f_{46}e^{-\la\tau_5}\\[1mm]
	0 & 0 & -c_{53}e^{-\la\tau_2}& 0 & \lambda -e_{55}e^{_\la\tau_4}-a_{55} & 0 \\[1mm]
	0 & 0 & 0& d_{64}e^{-\la\tau_5} & 0 & \lambda-f_{66}e^{-\la\tau_5}-a_{66} \\[1mm]
	
	\end{vmatrix} = 0
	$
}
\end{equation}
has only roots with negative real parts, the same will be true for the perturbed equation (8).
\\
Equation (9) factors into two distinct equations
\begin{equation}
(\la-a_{11})(\la-a_{22}-b_{22}e^{-\la \tau_1})+\eta^2x_1^*x_2^*e^{-\la \tau_1}=0
\end{equation}
and $d_1=0$ with 
\begin{equation}
\resizebox{\textwidth}{!}{%
	$
	d_1=\begin{vmatrix}
	
	 \lambda -a_{33}-c_{33}e^{-\la\tau_2} & 0 & -e_{35}e^{-\la\tau_4} & 0 \\[1mm]
	 0 & \lambda -a_{44}-d_{44}e^{_\la\tau_3} & 0 & -f_{46}e^{-\la\tau_5}\\[1mm]
	 -c_{53}e^{-\la\tau_2}& 0 & \lambda -e_{55}e^{_\la\tau_4}-a_{55} & 0 \\[1mm]
	 0& d_{64}e^{-\la\tau_5} & 0 & \lambda-f_{66}e^{-\la\tau_5}-a_{66} \\[1mm]
	
	\end{vmatrix} = 0
	$
}
\end{equation}
Replacing the values for the matrix coefficients, the equation (10) becomes
\begin{equation}
(\la+\mu)^3-\mu(\la^2+2 \mu\la+\mu^2-x_2^*)e^{-\la \tau_1}=0
\end{equation}
For $\tau_1=0$ one obtains the third degree polynomial 
$$\la^3+2\la^2\mu+\mu^2 \la+\mu x_2^*$$ that, according to the Routh-Hurwitz criterion, is stable if and only if $$2\mu^2>x_2^*.$$
Suppose this condition fulfilled. Then stability can be lost if and only if, when $\tau_1$ varies, some roots of the characteristic equation cross the imaginary axis from left to right.  So, suppose $\la=i\ome$ is a purely imaginary root of equation (10). Then, separating the real and imaginary parts in (11), the following system is obtained for $cos\ome\tau, sin\ome\tau$:
$$\begin{cases}
(\mu x_2^*-\mu^3+\mu \ome^2) cos\ome \tau-2 \mu^2 \ome sin\ome \tau=3 \mu\ome^2-\mu^3\\
-2 \mu^2\ome cos \ome \tau-(\mu x_2^*-\mu^3+\mu\ome^2) sin\ome\tau=\ome^3-3\mu^2\omega
\end{cases}$$
 For the equation (11) one can use the Schur complement formula and obtain the factorization into two equations
 \begin{equation}
 (\la-a_{33})(\la-a_{55})-(\la-a_{55})c_{33}e^{-\la\tau_2}-(\la-a_{33})e_{55}e^{-\la\tau_4}+c_{33}e_{55}e^{-\la(\tau_2+\tau_4)}
 \end{equation}
 and 
 \begin{equation}
 (\la-a_{44})(\la-a_{66})-(\la-a_{66})d_{44}e^{-\la\tau_3}-(\la-a_{44})f_{66}e^{-\la\tau_5}+d_{44}f_{66}e^{-\la(\tau_3+\tau_5)}
 \end{equation}
 
 Equations (13) and (14) can be studied following the approach in \cite{W}.
 \subsection{Extended model with antimicrobial resistance}
 To incorporate the effects of antimicrobial resistance, we extend the original model by adding two new variables:
 \begin{itemize}
 	\item $x_7$: Prevalence of antimicrobial resistance in the community (0-100\%)
 	\item $x_8$: Inappropriate antibiotic consumption rate
 \end{itemize}
 These new variables are governed by the following equations, which we append to the original system:
 \begin{empheq}[left=\empheqlbrace]{align}
 \dot{x}_{7} &= \alpha_R \cdot \frac{x_8}{K + x_8} \cdot \left(1 - \frac{x_7}{R_{max}}\right) - \delta_R \cdot x_7 \cdot (x_5 + x_6)
 \label{eq:7}\\[5pt]
 \dot{x}_{8} &= \beta_1 \cdot x_3 + \beta_2 \cdot x_4 - \gamma_1 \cdot x_5 - \gamma_2 \cdot x_6 - \mu_A \cdot x_8 + \sigma \cdot x_7 \cdot x_1
 \label{eq:8}
 \end{empheq}
 The parameters in the new equations are:
 \begin{itemize}
 	\item $\alpha_R$: Maximum rate of resistance development
 	\item $K$: Half-saturation constant for antibiotic effect on resistance
 	\item $R_{max}$: Theoretical maximum resistance level
 	\item $\delta_R$: Rate at which proper practices reduce resistance
 	\item $\beta_1$: Rate at which believers of fake news strain 1 contribute to inappropriate antibiotic use
 	\item $\beta_2$: Rate at which believers of fake news strain 2 contribute to inappropriate antibiotic use
 	\item $\gamma_1$: Rate at which skeptics of fake news strain 1 reduce inappropriate antibiotic use
 	\item $\gamma_2$: Rate at which skeptics of fake news strain 2 reduce inappropriate antibiotic use
 	\item $\mu_A$: Natural decay rate of inappropriate antibiotic use
 	\item $\sigma$: Feedback parameter: effect of resistance on antibiotic-seeking behavior among susceptible individuals
 \end{itemize}
 The biological meaning of these equations is as follows:
 
 \begin{itemize}
 	\item The first equation (Eq. \ref{eq:7}) models the development of antimicrobial resistance. Resistance increases with inappropriate antibiotic use ($x_8$) following saturation kinetics, with a logistic growth term $(1 - \frac{x_7}{R_{max}})$ that prevents unbounded growth. Resistance decreases when skeptics/fact-checkers ($x_5 + x_6$) practice proper antibiotic stewardship.
 	
 	\item The second equation (Eq. \ref{eq:8}) tracks inappropriate antibiotic consumption. This increases due to believers of fake news (both strains, with potentially different rates), decreases due to skeptics practicing proper use, naturally decays over time, and includes a feedback term where existing resistance can drive more antibiotic-seeking behavior among the susceptible population.
 \end{itemize}
\newpage
  \section{Numerical analysis and Matlab's simulations}
  To illustrate our theoretical results, we present numerical simulations using the following parameter values:
  \begin{table}[ht]
 \setlength{\tabcolsep}{4pt} 
 \centering
 \begin{tabular}{|p{2.5cm}|p{2cm}|p{2.5cm}|p{2cm}|}
 		\hline
 		Parameter & Value & Parameter & Value \\
 		\hline
 		$\Lambda$ & 100 & $\mu$ & 0.05 \\
 		\hline
 		$\eta$ & 0.20 & $\beta_3, \beta_4$ & 0.30, 0.20 \\
 		\hline
 		$\alpha_3, \alpha_4$ & 0.10, 0.10 & $\varepsilon_3, \varepsilon_4$ & 0.10, 0.10 \\
 		\hline
 		$\mu_3, \mu_4$ & 0.10, 0.10 & $\nu_3, \nu_4$ & 0.10, 0.10 \\
 		\hline
 		$\gamma_3, \gamma_4$ & 0.05, 0.05 & $\tau_1, \tau_2, \tau_3, \tau_4, \tau_5$ & 1, 1, 1, 1, 1 \\
 		\hline
 		$\alpha_R$ & 0.15 & $\delta_R$ & 0.10 \\
 		\hline
 		$\beta_1, \beta_2$ & 0.20, 0.15 & $\gamma_1, \gamma_2$ & 0.10, 0.08 \\
 		\hline
 		$\mu_A$ & 0.05 & $\sigma$ & 0.10 \\
 		\hline
 		$K$ & 10 & $R_{max}$ & 1 \\
 		\hline
 	\end{tabular}
 	\caption{Parameter values used in numerical simulations}
 \end{table}
\begin{figure}[H]
	\centering
	\includegraphics[width=\textwidth, keepaspectratio]{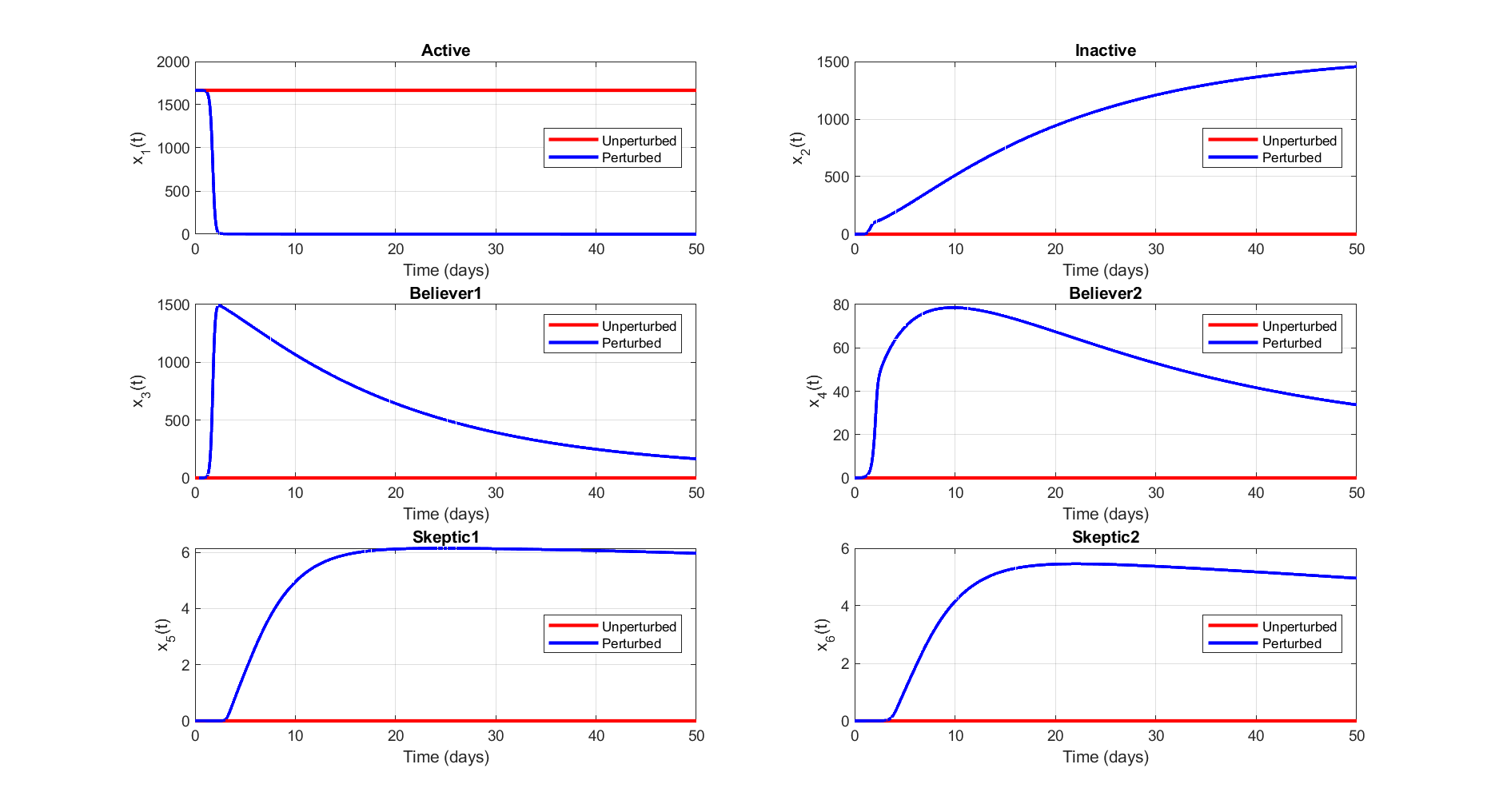}
	\caption{\(E_1= (1666.67,0,\dots,0)\) unstable under \(\tau_1=1,\dots,\tau_5=3\)}
\end{figure}

\begin{figure}[H] 
\centering
	\includegraphics[width=\textwidth, keepaspectratio]{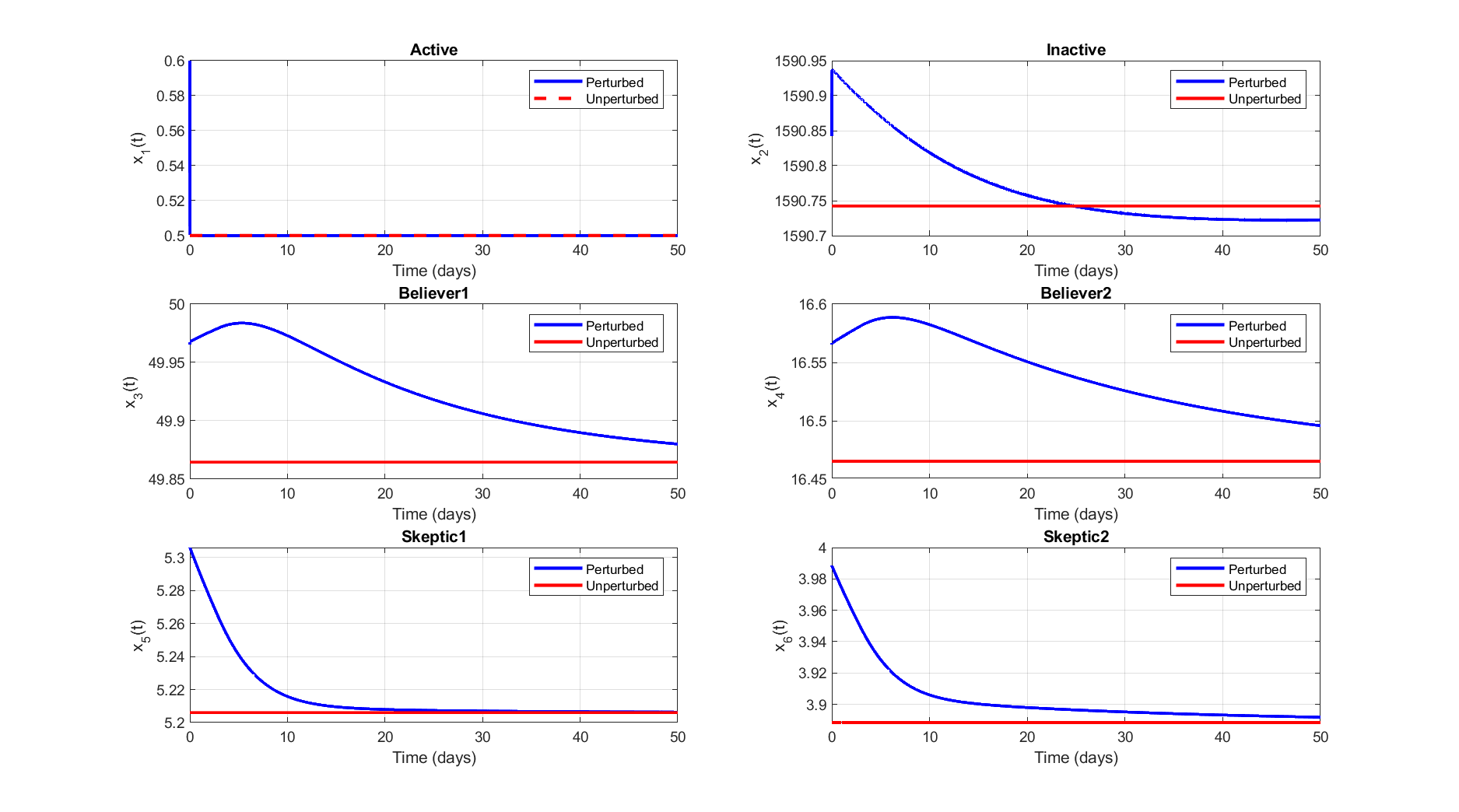}
\caption{\(E_2= (0.5,1590.742559,49.864465,16.465254,5.205975 3.888413)\) is stable \(\forall\tau\geq0\)}
\end{figure}
\begin{figure}[H] 
\centering
	\includegraphics[width=\textwidth, keepaspectratio]{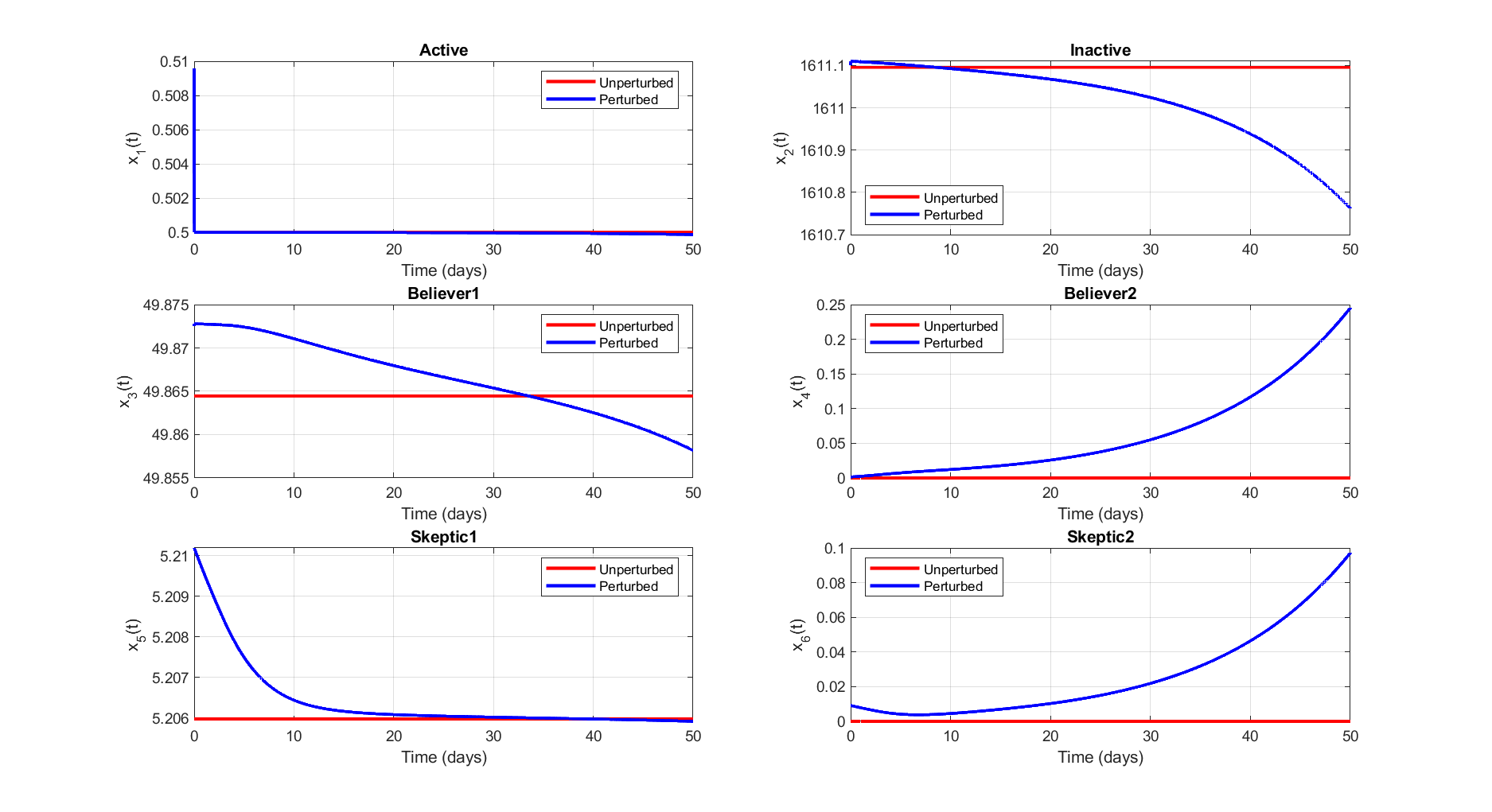}
\caption{Equilibrium point \(E_3= (0.5, 1611.096226,49.864465,0,5.205975, 0)\) is unstable under the delay values \(\tau_1=1, \tau_{2}=2, \tau_{3}=2, \tau_{4}=3, \tau_{5}=3\)}
\end{figure}
\begin{figure}[H] 
\centering
	\includegraphics[width=\textwidth, keepaspectratio]{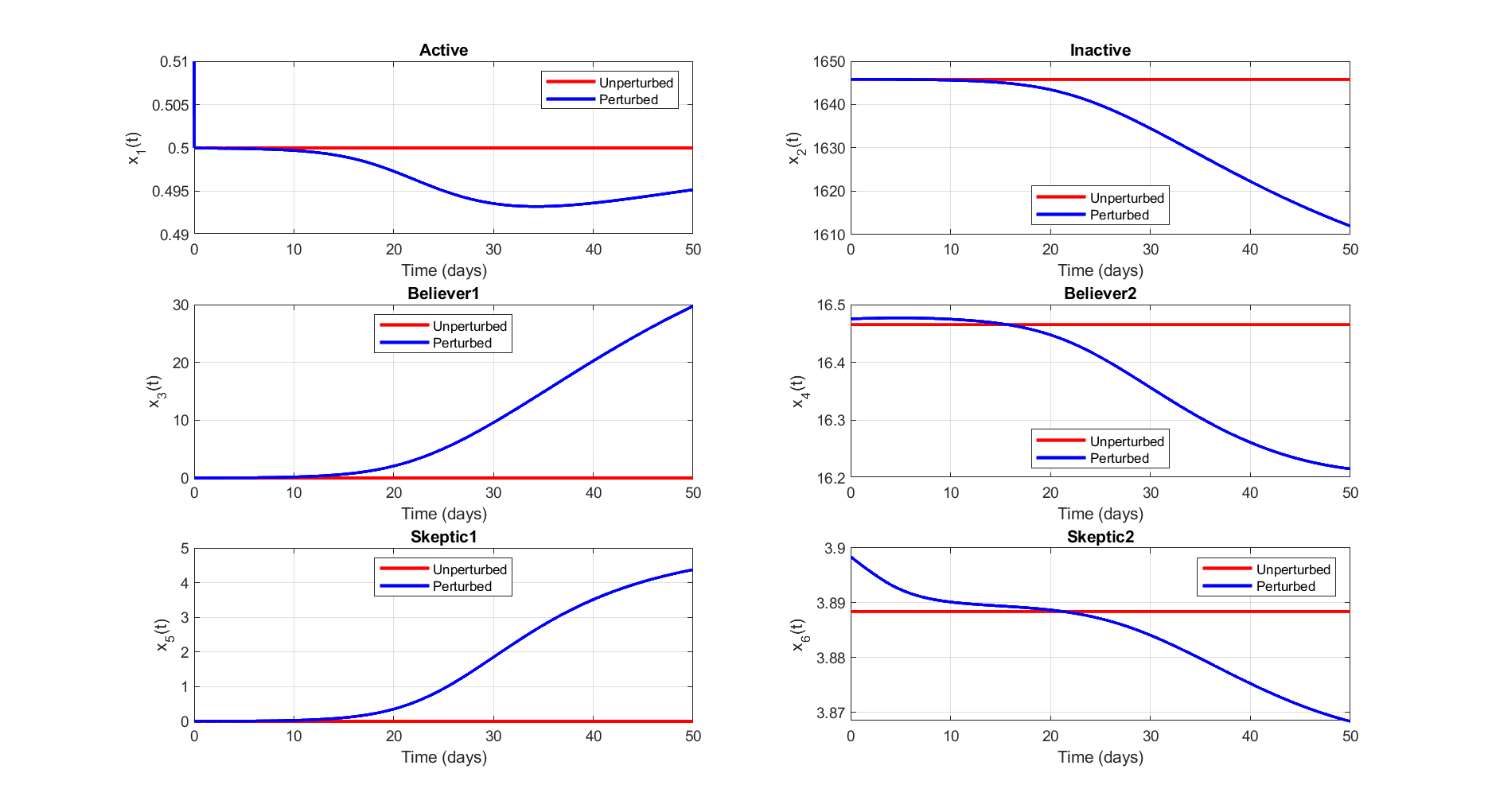}
\caption{\(E_4= (0.5, 1645.813,0,16.465254,0,3.888413)\) is unstable under the delay values \(\tau_1=1, \tau_{2}=2, \tau_{3}=2, \tau_{4}=3, \tau_{5}=3\)}
\end{figure}
\begin{figure}[H] 
\centering
	\includegraphics[width=\textwidth, keepaspectratio]{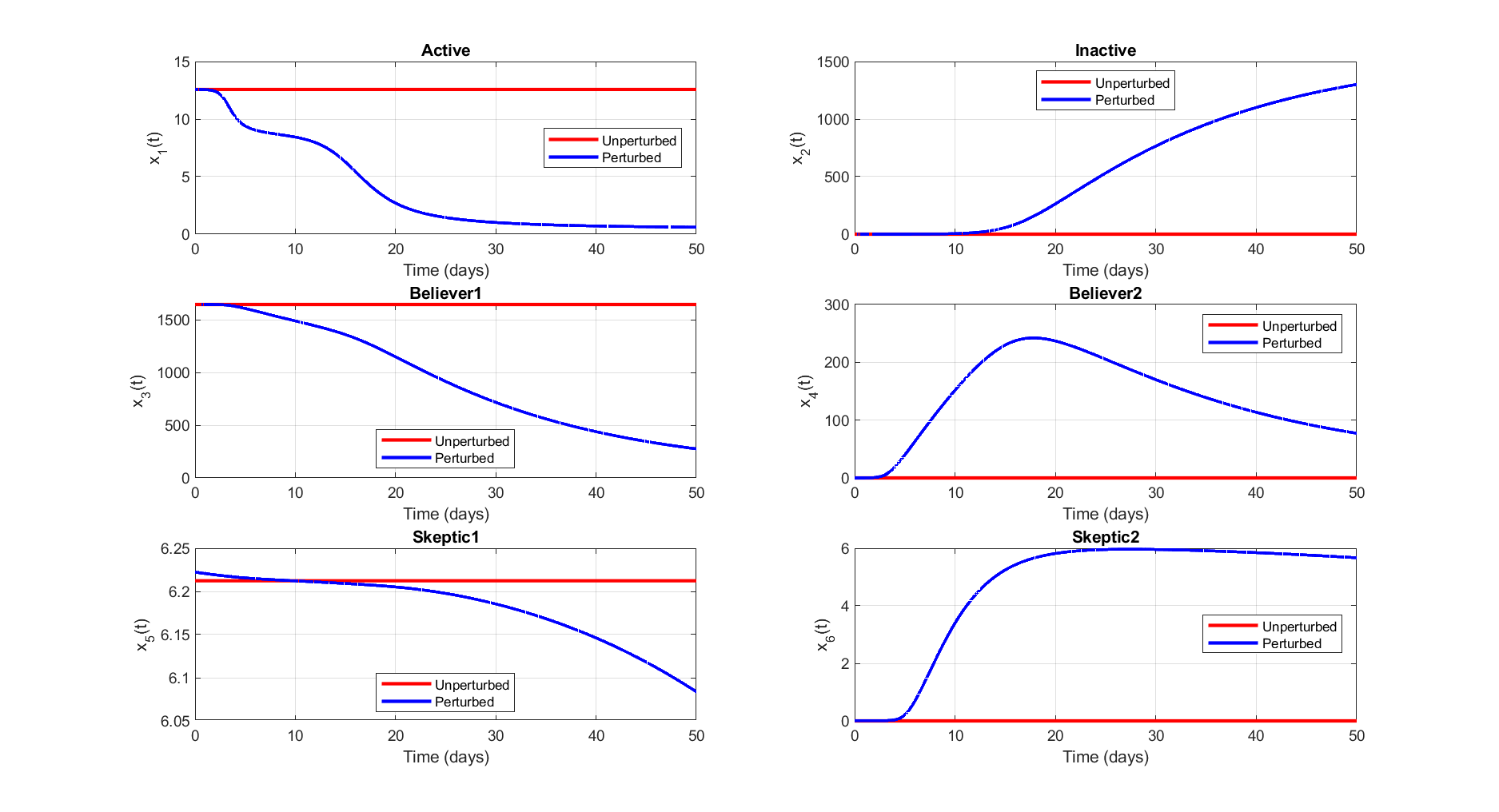}
\caption{\(E_5=( 12.575638,0,1647.878727,0,6.212301,0)\) is unstable under the delay values \(\tau_1=1, \tau_{2}=2, \tau_{3}=2, \tau_{4}=3, \tau_{5}=3\)}
\end{figure}
\begin{figure}[H] 
\centering
	\includegraphics[width=\textwidth, keepaspectratio]{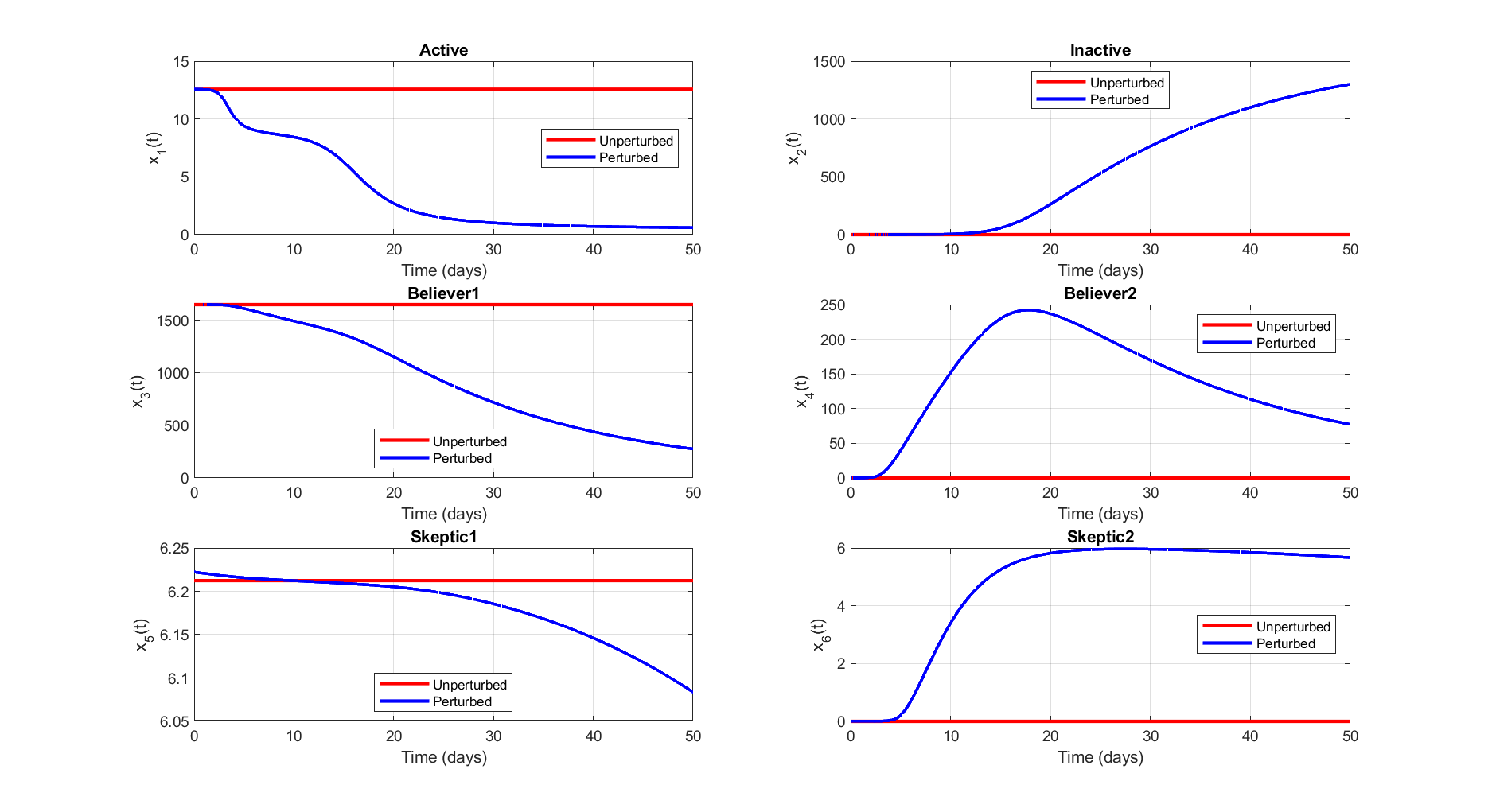}
\caption{ \(E_6=( 12.575638,0,1647.878727,0,6.212301,0)\) is unstable under the delay values \(\tau_1=1, \tau_{2}=2, \tau_{3}=2, \tau_{4}=3, \tau_{5}=3\)}
\end{figure}
\begin{figure}[H] 
\centering
	\includegraphics[width=\textwidth, keepaspectratio]{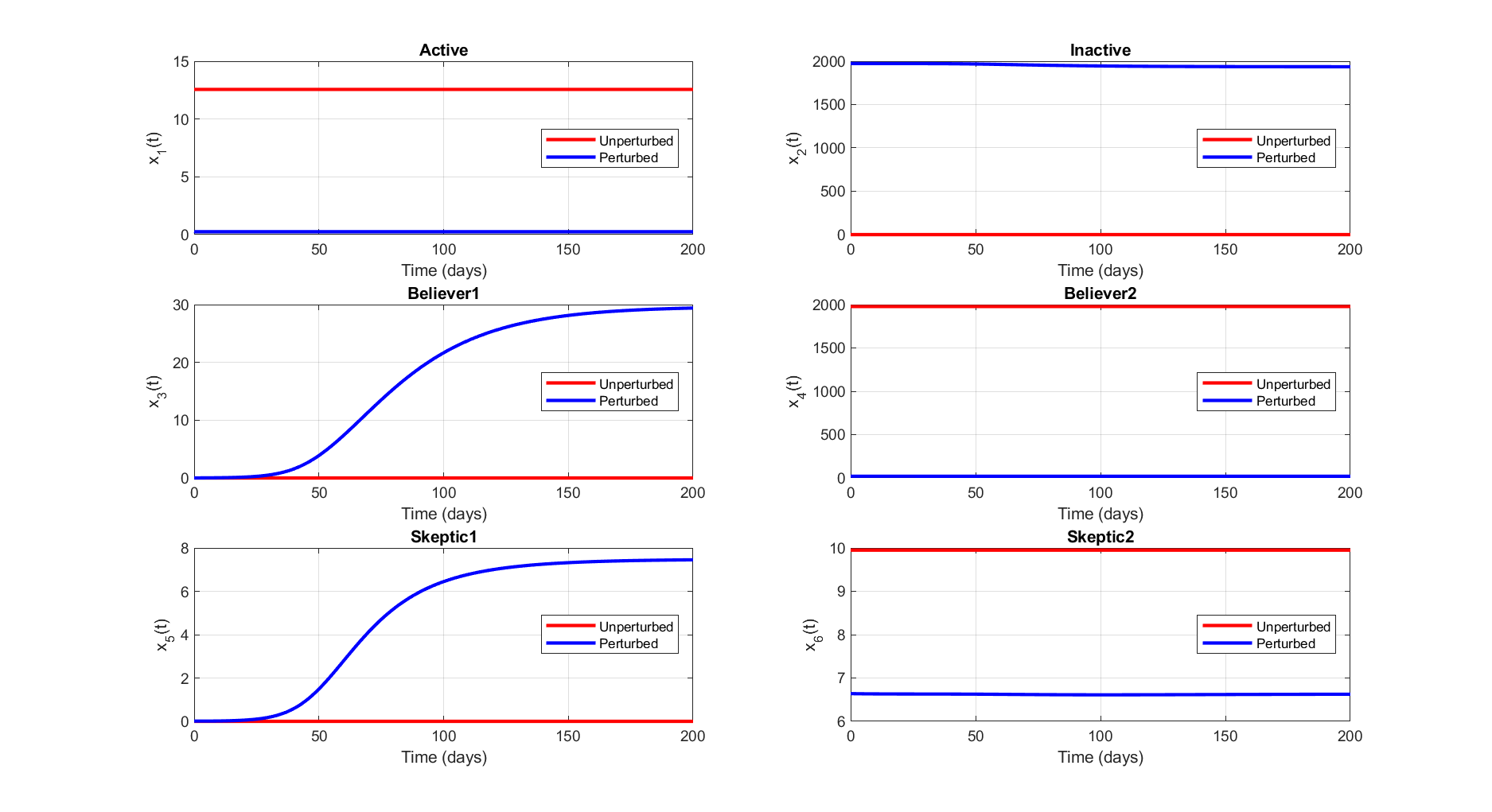}
\caption{ \(E_7=( 25.152003,0,0,1635.302651,0,6.212013)\) is unstable under the delay values  \(\tau_1=1, \tau_{2}=2, \tau_{3}=2, \tau_{4}=3, \tau_{5}=3\)}
\end{figure}
\section{Conclusion}
We’ve presented a model that shows how fake news spreads on social media, particularly when it’s related to antibiotic use. This new model connects the spread of false information with real health problems, like antibiotic resistance.

The model includes different kinds of misinformation, delays in how people respond, and how feedback works. It helps us understand how wrong ideas about antibiotics can lead to the growth of bacteria that are harder to treat.

Our findings show that even when there are efforts to fight misinformation, false beliefs can still stick around and cause serious issues. This shows how hard it is for public health experts to deal with both misinformation and antibiotic resistance at the same time.
\\
The model suggests that good strategies should focus on stopping the spread of misinformation, helping people shift to accurate beliefs, and addressing several false ideas at once. These insights can help guide better ways to reduce the impact of misinformation on the global fight against antibiotic resistance.
\bibliographystyle{plain}

\end{document}